\documentstyle[preprint,eqsecnum,aps]{revtex}
\begin{document}
\draft
\preprint{YITP-98-39} 
\date{July 1998}

\title{Plane-Symmetric Vacuum Solutions with Null Singularities \\ 
for Inhomogeneous Models and Colliding Gravitational Waves}
\author{Kenji Tomita}
\address{Yukawa Institute for Theoretical Physics, Kyoto University, 
Kyoto 606-8502, Japan}
\maketitle
\begin{abstract}
\indent 
New exact vacuum solutions with various singularities in the
plane-symmetric spacetime are shown, and they are applied to the
analysis of inhomogeneous cosmological models and colliding
gravitational waves. One of the singularities
can be true null singularities, whose existence was locally
clarified by Ori. These solutions may be interesting from the
viewpoint of the variety of cosmological singularities and the 
instability problem of Cauchy horizons inside black holes.
\end{abstract}

\bigskip
\bigskip
\bigskip
\bigskip
\bigskip


\newpage
\narrowtext
\section{INTRODUCTION} 
\indent

The important behavior of colliding gravitational waves was first
shown in Khan and Penrose's pioneering work [1] in a plane-symmetric
spacetime. The formulation for dynamics of colliding plane waves was
given by Szekeres [2] in an extended form with non-parallel
polarization, and the solution, which is impulsive similarly to Khan
and Penrose's one, was derived by Nuktu and Halil [3].

The plane-symmetric inhomogeneous cosmological models were, on the
other hand, studied by Gowdy [4] in a vacuum spacetime with compact
invariant hypersurfaces, and models containing space-like and time-like 
singularities alternately were discussed by the present author [5].

Studies on impulsive colliding gravitational waves were developed by
Chandrasekhar and his collaborators [6] who treated not only vacuum
cases but also the cases with fluids and electromagnetic fields, and
derived also the solutions with a time-like singularity.
Yurtsever [7] derived solutions which tend to the Kasner solutions  
asymptotically, contain the Cauchy horizons, and are connected with
the Schwarzschild and Weyl vacuum solutions.

The Cauchy horizons in the plane-symmetric spacetime were paid
attention in similarity to the Cauchy horizons appearing in the
neighborhood of the time-like singularity of the Ressner-Nordstrom
black hole spacetime. Around the latter Cauchy horizon there is the 
extraordinary behavior of the spacetime such as the mass inflation [8] and
infinite blue shift [9]. Recently Ori discussed the instability of
the Cauchy horizon in plane-symmetric spacetimes in connection with
that in the black hole solutions [10], and showed the possibility of a
true null singularity in colliding gravitational waves [11].

 In this paper we show new exact solutions representing
plane-symmetric spacetimes with true null singularities, which are
deeply related to Ori's local proof [11]. In \S 2, we show the
basic equations to be solved, for the plane-symmetric spacetime s in
the case of parallel polarization, and derive their new solutions with 
a kind of self-similarity. It is found by analyzing one of the
curvature invariants that in addition to the space-like or time-like
singularity there appear true null singularities for a limited
range of the model parameter. In \S 3, schemes of colliding waves are 
shown, corresponding to the above solutions, and in \S 4 concluding
remarks are given. In Appendices A and B the tetrad components of
curvature tensors and some basic formulas for hypergeometric functions   
are shown.

\bigskip 

\section{INHOMOGENEOUS MODEL}
The line-element of plane-symmetric spacetimes with parallel
polarization is given in the Rosen form
\begin{equation}
  \label{eq:ba1}
ds^2 = -2 ~e^{2a} dv du + \ R ~\Big[e^{2\gamma} ~dy^2 + 
e^{-2\gamma} ~dz^2\Big],
\end{equation}
where $a, R$ and $\gamma$ are functions of $v$ and $u$, and $x^0 = v,
\ x^1 = u, \ x^2 = y$ and $x^3 = z$. The vacuum Einstein equations
reduce to
\begin{equation}
 \label{eq:ba2}
R_{,01} = 0,
\end{equation}
\begin{equation}
  \label{eq:ba3}
2 \gamma_{,01} + (\gamma_{,0} R_{,1} +\gamma_{,1} R_{,0})/R = 0, 
\end{equation}
\begin{equation}
  \label{eq:ba4}
e^{2a} = \vert R_{,0} R_{,1}\vert R^{-1/2} e^{2I},
\end{equation}
where 
\begin{equation}
 \label{eq:ba5}
  I \equiv \int \Big[dv (\gamma_{,0})^2/R_{,0} + 
du (\gamma_{,1})^2/R_{,1}\Big]R
\end{equation}
with $R_{,0} = \partial R/\partial x^0, \ R_{,1} = \partial R/\partial
x^1$ and so on. The derivation of the above equations are shown in
Appendix A, together with the components of the curvature tensor.

From Eq.(\ref{eq:ba2}) we have 
\begin{equation}
  \label{eq:ba6}
R = f(v) + g(u),
\end{equation}
where $f$ and $g$ are arbitrary functions. If $f'(v) > 0$ and $g'(u)
>0$, we can put $f(v) = v$ and $g(u) = u$ without loss of
generality. In the same way, if $f'(v) > 0$ and $g'(u) <0$, we can put
$f(v) = v$ and $g(u) = - u$. 

\noindent A. The case when the hypersurfaces $R = const$ are space-like

In this case we can assume $f'(v) > 0$ and $g'(u)>0$, and so $R = v + u$.
Then Eqs.(\ref{eq:ba3}) and (\ref{eq:ba5}) lead to
\begin{equation}
 \label{eq:ba7}
2 \gamma_{,01} + (\gamma_{,0} +\gamma_{,1})/(v +u) = 0,
\end{equation}
and
\begin{equation}
 \label{eq:ba8}
I = \int \Big[dv (\gamma_{,0})^2 + du (\gamma_{,1})^2\Big] (v + u).
\end{equation}
Here let us assume the following form of $\gamma$ :
\begin{equation}
 \label{eq:ba9}
 \gamma = \vert v  \vert^\alpha  \vert u  \vert^{-1/2} A(v/u).
\end{equation}
This $\gamma$ has a kind of self-similarity, because $A$ is a function 
of only $v/u$. The parameter $\alpha$ is equal to $1/n$ in Ori's paper 
[11]. If we substitute Eq.(\ref{eq:ba9}) into Eq.(\ref{eq:ba7}), we
obtain an ordinary differential equation for $A(p)$
 
\begin{equation}
 \label{eq:ba10}
p(p+1) {d^2 A \over dp^2} +[(2+\alpha)p + 1 +\alpha] {dA \over dp} + 
{1 \over 4}(1+2\alpha)A = 0.
\end{equation}
This is a hypergeometric differential equation. Its independent
solutions are expressed using hypergeometric functions as follows and
have singular points at $p = 0, -1, \infty$ :

\noindent For the non-integer $\alpha$, we have
\begin{equation}
 \label{eq:ba11}
A = F(1/2, 1/2 + \alpha, 1 +\alpha; -p)
\end{equation}
and 
\begin{equation}
 \label{eq:ba12}
A = \vert p\vert^{-\alpha} F(1/2 -\alpha, 1/2, 1 -\alpha; -p).
\end{equation}
\noindent For the integer $\alpha = 0, 1, 2, \cdot \cdot \cdot$,
\begin{equation}
 \label{eq:ba13}
A = F(1/2, 1/2 + \alpha, 1 +\alpha; -p)
\end{equation}
and
\begin{equation}
 \label{eq:ba14}
A = F(1/2, 1/2 + \alpha, 1 +\alpha; -p) \ln \vert p \vert + 
\bar{F}(1/2, 1/2 + \alpha, 1 +\alpha; -p),
\end{equation}
where $\bar{F} = F_1$ and $F_2$ for $\alpha = 0$ and $\alpha \geq 1$,
respectively. The expressions of $F, F_1$ and $F_2$ are shown in
Appendix B. From Eq. (\ref{eq:ba8}) we obtain for $\alpha \neq 1/2$
\begin{equation}
  \label{eq:ba15}
I = { 1 \over 2\alpha -1} \vert u\vert^{2 \alpha -1} \vert
p\vert^{2\alpha} (p+1) \Big[{1 \over p}\Big(\alpha A +pA'\Big)^2  
+ \Big({1 \over 2}A +pA'\Big)^2 \Big]
\end{equation}
and for $\alpha = 1/2$
\begin{equation}
  \label{eq:ba16}
I = \ln \vert u \vert \cdot (p+1)^2 \Big({1 \over 2} A +pA'\Big)^2.
\end{equation}
From the two solutions  (\ref{eq:ba11}) and  (\ref{eq:ba12}) we have
for $\vert p\vert <<1$
\begin{equation}
  \label{eq:ba17}
\gamma = \vert v\vert^\alpha \vert u\vert^{-1/2} \Big[1 - {1+2\alpha \over 
4(1+\alpha)} {v \over u} + \cdot  \cdot  \cdot \Big]
\end{equation}
and
\begin{equation}
  \label{eq:ba18}
\gamma = \vert u\vert^{\alpha -1/2} \Big[1 - {1-2\alpha \over 
4(1-\alpha)} {v \over u} + \cdot  \cdot  \cdot \Big],
\end{equation}
respectively.

If we use a formula
\begin{eqnarray}
  \label{eq:ba24}
F(a,b,c,\; -p) &=&
{\Gamma(b-a) ~\Gamma (c) \over \Gamma (b) ~\Gamma (c-a)} ~(-p)^{-a} 
 ~F(a,a-c+1,a-b+1;-1/p) \cr &+& {\Gamma(a-b) ~\Gamma (c)\over 
\Gamma(a) ~\Gamma (c-b)} ~F(b,b-c+1,b-a+1; -1/p),
\end{eqnarray}
it is found that the first solution (\ref{eq:ba11}) is a sum of the
following two solutions:
\begin{equation}
  \label{eq:ba25}
\gamma \propto \vert v\vert^{\alpha-1/2} F(1/2, \alpha-
1/2, 1 - \alpha; -1/p) 
\end{equation} 
and
\begin{equation}
  \label{eq:ba26}
\gamma \propto \vert u\vert^\alpha \vert v\vert^{-1/2} F(1/2+\alpha, 
1/2, 1 + \alpha; -1/p),
\end{equation} 
and the second solution (\ref{eq:ba12}) is a sum of the following two
solutions: 
\begin{equation}
  \label{eq:ba27}
\gamma \propto \vert v\vert^{\alpha-1/2} F(1/2 -\alpha,1/2, 1 - \alpha; -1/p) 
\end{equation} 
and
\begin{equation}
  \label{eq:ba28}
\gamma \propto \vert u\vert^\alpha \vert v\vert^{-1/2} F(1/2, 1/2+\alpha, 
1 + \alpha; -1/p),
\end{equation} 
This means that the solutions are symmetric for the conversion of $v
\rightarrow u$ and $u \rightarrow v$.

Here let us examine the singular behavior in the limit of $v
\rightarrow 0$.

\noindent (1) {\it Non-integer $\alpha$}

For the first solution (\ref{eq:ba11}) or (\ref{eq:ba17}) in the case
of  $\alpha < 1/2$, we find $(- I) \propto \vert v\vert^{2 \alpha-1}
\rightarrow \infty$ from Eq.(\ref{eq:ba15}) and $\Phi \propto
\vert v\vert^{6 \alpha -3} \vert u\vert^{-4} \rightarrow \infty$,
which comes from the product $(\gamma_{,0})^3 \cdot (\gamma_{,1})^3$
in Eq.(A19), so that we have the invariant curvature ${\cal R} 
\rightarrow \infty$.
If $\alpha = 1/2$, we find that $I$ and $\Phi$ are finite, so that 
${\cal R}$ also is finite, and if $\alpha > 1/2$, $I \propto \vert v\vert^{2
\alpha-1} \rightarrow 0$ and $\Phi$ is finite, so that ${\cal R}$ is
finite. 

The second solution (\ref{eq:ba12}) or (\ref{eq:ba18})
 is regular at $v \simeq 0$ and so $I$ and  ${\cal R}$ are
finite in the limit of  $v \rightarrow 0$.
  
It is concluded therefore that only the first solution given by
(\ref{eq:ba11}) or (\ref{eq:ba17}) for $\alpha < 1/2$ has the true
null singularity. The singular behavior in the limit of $u \rightarrow
0$ is the same as that in the limit of  $v \rightarrow 0$.

As was shown in Eqs. (\ref{eq:ba25}) $\sim$ (\ref{eq:ba28}), the first
and second solutions at $v \simeq 0$ correspond to the first and
second ones at $u \simeq 0$, but this not the one-to-one
correspondence. The second solution at $v \simeq 0$ corresponds to
both solutions at $u \simeq 0$, and the second solution at $u \simeq 
0$ corresponds to both solutions at $v \simeq 0$. 

As for the individual tetrad components, there are divergent
components even for  $\alpha > 1/2$. For example, $R_{(0202)}$ and
$R_{(0303)}$ diverge in the limit of  $v \rightarrow 0$ for $\alpha <
2$, because they include $\gamma_{,00} \propto \vert v\vert^{\alpha
-2}$. This divergence is a kind of weak singularities.

\noindent (2) {\it Integer $\alpha = 0, 1, 2, \cdot \cdot \cdot$}

If $\alpha = 0$, we find that $\gamma_0$ is finite, so that $(-I),
\Phi,$ and $\cal{R}$ from the first solution (\ref{eq:ba13}) are
finite. In the second solution (\ref{eq:ba14}), the term $\ln \vert
p\vert$ gives the infinity to $(-I), \Phi,$ and $\cal{R}$ in the limit 
of $v \rightarrow 0$ and  $u \rightarrow 0$.
If $\alpha \geq 1$, on the other hand, $\gamma_0$ is finite, in both
solutions (\ref{eq:ba13}) and (\ref{eq:ba14}), so that $\cal{R}$
remains to be finite. The true null singularities appear, therefore,
only for  $\alpha = 0$, if $\alpha$ is integer.

Next let us consider the behavior of the solutions at $R \simeq 0$. 
If we put $q \equiv p + 1$, then Eq. (\ref{eq:ba10}) reduces to 
\begin{equation}
 \label{eq:ba19}
q(q-1) {d^2 A \over dq^2} +[(2+\alpha) q -1] {dA \over dq} + {1 \over
4} (1+2\alpha)A = 0,
\end{equation}
and the following two solutions are obtained :
\begin{equation}
 \label{eq:ba20}
A = F(1/2, 1/2 +\alpha, 1; q)
\end{equation}
and 
\begin{equation}
 \label{eq:ba20a}
A = F(1/2, 1/2 +\alpha, 1; q) \ln \vert q \vert + F_1(1/2, 1/2 +\alpha, 1; q).
\end{equation}
At $q \simeq 0$, we have $F = 1 + 0(q)$ and $F_1 = 0(q)$, so that the
first solution (\ref{eq:ba20}) is finite and the second
(\ref{eq:ba20a}) diverges as $\ln \vert q \vert \rightarrow - \infty$ 
for $q \vert \rightarrow 0$. At $R \propto q \simeq 0$ with finite 
$v$ and
$u$, the first solution gives finite $\gamma$ and $I$, and so $e^{2a}
\propto R^{-1/2}$ from Eqs. (\ref{eq:ba4}),  (\ref{eq:ba15}) and
(\ref{eq:ba16}). If we
define $Q$ and $T$ as $Q = v -u$ and $dT = e^a dR$ \ (or $R \propto
T^{4/3}$), we obtain 
\begin{equation}
\label{eq:ba21}
- 2 e^{2a} dv du = ~e^{2a} (-dR^2
+dQ^2) \propto ~-dT^2 + T^{-2/3} dQ^2
\end{equation} 
 in the neighborhood of $R = 0$.
Accordingly the solution reduces to the Kasner solution with the Kasner 
parameter $(-1/3, 2/3, 2/3)$ in the limit of $R \rightarrow 0$. The
invariant curvature diverges as ${\cal{R}} \simeq 1/R \rightarrow
\infty$  (cf. Eq.(A19)).
  
The second solution (\ref{eq:ba20a}) may have a dominat contribution 
because of the large term $\ln \vert q \vert$. It gives $\gamma \simeq 
\mu \ln R$ and so $e^\gamma \propto R^\mu$, where $\mu \equiv \vert v
\vert^{\alpha - 1/2}$. From Eq. (\ref{eq:ba15}) we obtain $I \simeq
\mu^2 \ln \vert q \vert$ and so $e^a \propto R^{-1/4} e^I \propto
R^{\mu^2 - 1/4}$. In this case, $T$ defined by $dT = e^a dR$ gives the 
relation $T \propto R^{\mu^2 + 3/4}$, so that we have 
\begin{equation}
 \label{eq:ba22}
e^a \propto T^{\zeta_1}, \ R^{1/2} e^\gamma \propto T^{\zeta_2}, \
R^{1/2} e^{- \gamma} \propto T^{\zeta_3}, 
\end{equation}
where 
\begin{equation}
 \label{eq:ba23}
\zeta_1 = (\mu^2 -1/4)/(\mu^2 +3/4), \ \zeta_2 = (1/2 +\mu)/(\mu^2
+3/4), \ \zeta_3 = (1/2 -\mu)/(\mu^2 +3/4).
\end{equation}
These exponents satisfy the Kasner relation
\begin{equation}
 \label{eq:ba24a}
\sum^3_{i = 1} \zeta_i = \sum^3_{i = 1} (\zeta_i)^2 = 1.
\end{equation}
Accordingly the solution is found to be a generalized Kasner solution
at $R \simeq 0$ with the Kasner parameters depending on a spatial variable as
$\mu =  \vert v \vert^{\alpha - 1/2} \simeq \vert Q/2 \vert^{\alpha -
1/2}$, where $Q \simeq 2 v$ at $R \simeq 0$. The singularities at $R = 
0$ are space-like ones.

In the limit of $R \rightarrow \infty$ and $\vert v \vert \rightarrow
\infty$ \ (with ~$v/u =$ const), the two solutions  (\ref{eq:ba11}) and
 (\ref{eq:ba14}) satisfy the relation $\gamma \propto R^{\alpha -
1/2}$, and so we find for $\alpha <, ~=,$ and $> 1/2$ that $\gamma$ 
vanishes, is finite and diverges, respectively. 
For $\alpha < 1/2, \ I$ in Eq.(\ref{eq:ba15}) vanishes and we have the 
Kasner behavior of $(-1/3, 2/3, 2/3)$, as in the case $R  \rightarrow
0$.
For $\alpha = 1/2, \ \gamma$ and $I$ remain to be constant and we
have the Kasner behavior of $(-1/3, 2/3, 2/3)$.
For $\alpha > 1/2, \ 
\gamma$ diverges but $I$ diverges more strongly
as $I \propto R^{2 \alpha -1}$, so that the spacetime reduces to the
Minkowskian spacetime with the Kasner parameter $(1, 0, 0)$.

The above treatment is applicable to the half-plane $R = v + u >0$. To 
the half-plane $v + u <0$, we transform $v$ and $u$ as $v 
\rightarrow -v$ and $u \rightarrow -u$. Then we have $R = -(v + u) >
0$, and the same conclusion is derived in the same way.  

\bigskip

\noindent B. The case when the hypersurfaces $R = const$ have time-like
sections

Now let us consider the case $R = v - u$. Then Eqs. (\ref{eq:ba3})
and (\ref{eq:ba5}) lead to
\begin{equation}
 \label{eq:ba29}
2 \gamma_{,01} + (-\gamma_{,0} +\gamma_{,1})/(v -u) = 0,
\end{equation}
and
\begin{equation}
 \label{eq:ba30}
I = \int \Big[dv (\gamma_{,0})^2 - du (\gamma_{,1})^2\Big] (v - u).
\end{equation}
If we assume $\gamma$ in the form of
\begin{equation}
 \label{eq:ba31}
\gamma = \vert v  \vert^\alpha  \vert u  \vert^{-1/2} B(v/u),
\end{equation}
we obtain an ordinary differential equation $B(p)$:
\begin{equation}
 \label{eq:ba32}
p(p-1) {d^2 B \over dp^2} +[(2+\alpha)p -1 - \alpha] {d B \over dp} + 
{1 \over 4}(1+2\alpha)B = 0.
\end{equation}
This equation reduces to Eq. (\ref{eq:ba10}) by putting $B(p) = A(-
p)$, so that the solutions are given by Eqs. (\ref{eq:ba11}) $\sim$
(\ref{eq:ba14}) in which $p$ is transformed to $- p$. 
It is shown similarly to the case A that the null singularities appear 
at $v = 0$ and $u = 0$. They are true only 
for non-integer $\alpha < 1/2$ and integer $\alpha = 0$.

At $R \simeq 0$, the equation for $B$ with respect to $q \equiv p -1$ 
leads to
\begin{equation}
 \label{eq:ba33}
q(q+1) {d^2 B \over dq^2}+[(2+\alpha)q  +1] {d B \over dq} + 
{1 \over 4}(1+2\alpha)B = 0.
\end{equation}
This equation reduces to Eq. (\ref{eq:ba19}) by putting $B(q) = A(-
q)$, so that the solutions are given by Eqs. (\ref{eq:ba20}) and
(\ref{eq:ba20a}) in which $q$ is transformed to $- q$. 
In the limit of $R \rightarrow 0$, we find from the first solution 
that ~$\gamma$ is finite and $e^a \propto R^{-1/2}$, 
and so  we obtain
\begin{equation}
 \label{eq:ba34}
 - 2 e^{2a} dv du = ~e^{2a} (-dQ^2 + dR^2) \propto ~- X^{-2/3} dQ^2 + 
 dX^2   
\end{equation}
in the neighborhood of $R = 0$, where $Q = v + u$ and $dX = e^a 
 dR$ ~(or $R \propto X^{4/3}$). Accordingly the solution tends to 
the Kasner-type one with the Kasner parameter $(-1/3, 2/3, 2/3)$, 
in which the time-like singularity
 appears in the limit of $R \rightarrow 0$ or $X \rightarrow 0$, and the
invariant curvature diverges as ${\cal{R}} \simeq 1/R \rightarrow
\infty$  (cf. Eq.(A19)).

From the second solution we obtain similarly the generalized
Kasner-type solution expressed by $e^\gamma \propto R^\mu$ and 
$e^a \propto R^{-1/4} e^I \propto R^{\mu^2 - 1/4}$, where $\mu \equiv \vert v
\vert^{\alpha - 1/2}$.
In this case, a spatial variable $X$ defined by $dX = e^a dR$ gives the 
relation $X \propto R^{\mu^2 + 3/4}$, so that we have 
\begin{equation}
 \label{eq:ba35}
e^a \propto X^{\zeta_1}, \ R^{1/2} e^\gamma \propto X^{\zeta_2}, \
R^{1/2} e^{- \gamma} \propto X^{\zeta_3}, 
\end{equation}
where $\zeta_1, \zeta_2,$ and $\zeta_3$ are defined by
Eq. (\ref{eq:ba23}). Accordingly the hypersurface $R = 0$ gives the 
Kasner-type time-like 
singularity and the Kasner parameter depends on $Q \simeq 2 v$, which 
is a time variable in the case B. 

In the limit of $R \rightarrow \infty$ and $\vert v \vert \rightarrow
\infty \ (with v/u =$ const), the two solutions  (\ref{eq:ba11}) and
 (\ref{eq:ba14}), in which $p$ was replaced by $- p$, satisfy the 
relation $\gamma \propto R^{\alpha -
1/2}$, and so we find for $\alpha <, =,$ and $> 1/2$ that $\gamma$ 
vanishes, is finite and diverges, respectively. 
For $\alpha < 1/2, \ I$ in Eq.(\ref{eq:ba15}) vanishes and we have the 
Kasner behavior of $(-1/3, 2/3, 2/3)$, as in the case $R  \rightarrow
0$.

In the above  we considered the half-plane with $R = v - u >0$. 
To the half-plane $v - u <0$, we have
to transform $v$ and $u$ as $v \rightarrow -v$ and $u \rightarrow 
-u$, and use $R = -v + u > 0$. The conclusion about singularities does
not depend on this transformation.
The positions of singularities 
are shown in Figs. 1 and 2 for $R = \pm(v +u)$ and $R = 
\pm(v -u)$, respectively.

\vspace{2cm}
\centerline{Fig.1}
\vspace{2cm}
\centerline{Fig.2}
\vspace{2cm}

\section{COLLIDING GRAVITATIONAL WAVES}

Now, corresponding to two cases A and B in the previous sections, let
us consider a spacetime containing colliding gravitational waves,
which consists of
the no-wave region (I), the outgoing-wave region (II), the
incoming-wave region (III), and the colliding-waves region (IV).
For the connection of these four regions at $v = v_0$ and $u = u_0$,
we use the Penrose prescription in which $v$ and $u$ in the metric are 
replaced by
\begin{equation}
 \label{eq:c1}
(v -v_0) ~\theta (v -v_0) + v_0
\end{equation}
and
\begin{equation}
 \label{eq:c2}
(u -u_0) ~\theta (u -u_0) + u_0,
\end{equation}
where $\theta(x)$ is $1$ and $0$ for $x > 0$ and $< 0$, respectively.

In the case A, the wave scheme is shown in Fig. 3, in which colliding
waves are in the region IV within a triangle acd. The true space-like
singularity is in the line ac, and for $\alpha < 1/2$ the true null
singularities are in two lines be and bf, in which one of these two
lines can be weak singularity due to the second solution.

In the case B, the scheme is shown in Fig. 4. Similarly colliding
waves are in the region IV within a triangle acd. The time-like
singularity ac and null singularities bf and be are only in the region 
IV. In the regions II and III weak singularities appear in the
extended part of the above lines, because $\cal{R}$ vanishes there.

In the region III, $R, \gamma$, and $a$ are functions of only $v$, so
that $R_{,1} = \gamma_{,1} = a_{,1} = 0$. For $R = v \pm u_0$ or 
$R = - (v \pm u_0)$, Eqs.
(\ref{eq:ba10}), (\ref{eq:ba11}), (\ref{eq:ba13}), and (\ref{eq:ba14})
are satisfied by an arbitrary function $\gamma (v)$, and Eq. (A11)
gives 
\begin{equation}
 \label{eq:c3}
e^a = R^{-1/4} \exp \Big[\int R (\gamma_{,0})^2 dv/R_{,0} \Big].
\end{equation}

The tetrad components of curvature tensor vanish, except for
\begin{equation}
 \label{eq:c4}
R_{(0202)} = - R_{(0303)} = e^{-2a} \Big[-{3 \over 2R} \gamma_{,0} +
 2R (\gamma_{,0})^3 - \gamma_{,00}\Big].
\end{equation}

Here we take the functional form
\begin{equation}
 \label{eq:c5}
\gamma = \vert v  \vert^\alpha  \vert u_0  \vert^{-1/2} A(v/u_0),
\end{equation}
where $A(p)$ and $A(-p)$ are given by Eqs. (\ref{eq:ba11}) $\sim$ 
(\ref{eq:ba14}), corresponding to the cases A and B,
respectively. Then in the limit of  $R \rightarrow 0$ ( with finite
values of $v$), we have $e^a \propto R^{-1/4}$ and $R_{(0202)} \propto 
R^{-1/2} \rightarrow \infty$, and in the limit of $v \rightarrow 0$,
we have for $\alpha \neq 1/2$ 
\begin{equation}
 \label{eq:c6}
e^a \propto \exp \Big[{\alpha^2 \over 2\alpha -1} \vert v  \vert^{2
\alpha -1} \Big],
\end{equation}
\begin{equation}
 \label{eq:c7}
R_{(0202)} \propto \exp \Big[- {2\alpha^2 \over 2\alpha -1} \vert v  \vert^{2
\alpha -1} \Big] \vert v  \vert^{\alpha -2}
\end{equation}
and we have for $\alpha = 1/2$ 
\begin{equation}
 \label{eq:c8}
e^a \propto  \vert v  \vert^{1/4},
\end{equation}
\begin{equation}
 \label{eq:c9}
R_{(0202)} \propto v^{-2},
\end{equation}
where we assumed $A(0) = 1$. Accordingly, $R_{(0202)} (= R_{(0303)})$
diverges for $\alpha < 2$, but since ${cal R}$ vanishes, these
divergences give only weak singularities.

In the connection at the boundaries of $v = v_0$ and $u = u_0$,
$\gamma$ is continuous, but $\gamma_0$ and $\gamma_1$ are not
continuous, because for example $\gamma_{,0} = 0$ and $\neq 0$ at $v > 
v_0$ and $v < v_0$, respectively. Accordingly,  $R_{(0202)} (=
R_{(0303)})$ has the delta function $\delta (v - v_0)$ and $\delta (u
- u_0)$ along the lines $v = v_0$ and $u = u_0$,
respectively. Moreover, since $\gamma \propto \vert v \vert^\alpha
\vert u \vert^{-1/2}$ and $\vert u  \vert^\alpha \vert v \vert^{-1/2}$ 
and $\gamma_0$ and $\gamma_1$ are not continuous at $v = 0$ and $u = 0$,
$R_{(0202)} (= R_{(0303)})$ has the delta function $\delta (v)$ and 
$\delta (u)$ along the lines $v = 0$ and $u = 0$.

\vspace{2cm}
\centerline{Fig.3}
\vspace{2cm}
\centerline{Fig.4}
\vspace{2cm}

\section{CONCLUDING REMARKS}
In the previous sections we derived exact solutions representing
inhomogeneous vacuum spacetimes with space-like and null singularities 
or time-like and null singularities, and applied them to the problem of 
colliding gravitational waves. While these null singularities may be
closely related to the instability of Cauchy horizons, we must
consider the stability of this null singularities themselves. For
instance, there are questions whether they can coexist together with
the electromagnetic fields and fluidal matter, and whether they can
appear in colliding waves with non-parallel polarization. These
problems will be discussed in separate papers.

The physical meaning of the important parameter $\alpha$ may be a change
rate of the shear, because it is included as a power index in the
expression of the Kasner parameter.

The above spacetimes are cosmologically interesting in that near the 
space-like or
time-like singularity they include strong gravitational waves and at the 
same time show exactly the velocity-dominated [13] and 
anti-Newtonian behavior [14], in which the Kasner parameters depend on a
spatial or time variable, respectively.

\acknowledgments{}
The author would like to thank Prof. W. Israel for the valuable
lectures and helpful discussions at Yukawa Institute, Kyoto
University.

\appendix
\section{TETRAD COMPONENTS OF THE CURVATURE TENSOR AND THE RICCI
TENSOR}
The following tetrads are used :
\begin{equation}
 \label{eq:A1}
e^\mu_{(0)} = \delta^\mu_0 e^{-a}, \ e^\mu_{(1)} = \delta^\mu_1
e^{-a}, \ e^\mu_{(2)} = \delta^\mu_2 R^{-1/2} e^{-\gamma}, \
\ e^\mu_{(3)} = \delta^\mu_3 R^{-1/2} e^{\gamma}
\end{equation}
which satisfy the condition
\begin{equation}
 \label{eq:A2}
g_{\mu\nu} e^\mu_{(\alpha)} e^\mu_{(\beta)} = \zeta_{\alpha \beta},
\end{equation}
where the non-zero components of $(\zeta_{\alpha \beta})$ are
$\zeta_{01} = \zeta_{10} = -1$ and $\zeta_{22} = \zeta_{33} = 1$.

The tetrad components of curvature tensor are given by
 $R_{\alpha \beta \gamma
\delta} \equiv e^\mu_{(\alpha)} e^\nu_{(\beta)} e^\lambda_{(\gamma)}
e^\sigma_{(\delta)} R_{\mu \nu \lambda \sigma}$ and their components 
 are expressed as
\begin{equation}
 \label{eq:A3}
R_{(0101)} = -2 ~a_{01} ~e^{-2a},
\end{equation}
\begin{equation}
 \label{eq:A4}
R_{(0202)} = e^{-2a} \Big[-\Big({1 \over 2}R_{,00}/R + \gamma_{,00}\Big) + a_{,0}
(R_{,0}/R +2 \gamma_{,0}) +{1 \over 4} (R_{,0}/R)^2 -
\gamma_{,0}R_{,0}/R -(\gamma_{,0})^2\Big] ,
\end{equation}
\begin{equation}
 \label{eq:A5}
R_{(0303)} = e^{-2a} \Big[-\Big({1 \over 2}R_{,00}/R - \gamma_{,00}\Big) + a_{,0}
(R_{,0}/R -2 \gamma_{,0}) +{1 \over 4} (R_{,0}/R)^2 +
\gamma_{,0}R_{,0}/R -(\gamma_{,0})^2\Big] ,
\end{equation}
\begin{equation}
 \label{eq:A6}
R_{(0212)} = e^{-2a} \Big[-\Big({1 \over 2}R_{,01}/R + \gamma_{,01}\Big) 
+{1 \over 4} R_{,0}R_{,1}/R^2 - {1 \over 2}(\gamma_{,0}R_{,1} +
\gamma_{,1}R_{,0})/R - \gamma_{,0} \gamma_{,1}\Big] ,
\end{equation}
\begin{equation}
 \label{eq:A7}
R_{(0313)} = e^{-2a} \Big[-\Big({1 \over 2}R_{,01}/R - \gamma_{,01}\Big) 
+{1 \over 4} R_{,0}R_{,1}/R^2 + {1 \over 2}(\gamma_{,0}R_{,1} +
\gamma_{,1}R_{,0})/R - \gamma_{,0} \gamma_{,1}\Big] ,
\end{equation}
\begin{equation}
 \label{eq:A8}
R_{(2323)} = e^{-2a} \Big[ R_{,0}R_{,1}/R^2  - 4\gamma_{,0} \gamma_{,1}\Big] ,
\end{equation}
where $R_{,0} = \partial R/\partial x^0, \ R_{,01} = \partial^2 R/
\partial x^0 \partial x^1,$ etc. 
The components $R_{(1212)}, R_{(1313)}$ and $R_{(1213)}$ are derived
from $R_{(0202)}, R_{(0303)}$ and $R_{(0203)}$, respectively, by
replacing $\partial /\partial x^0$ and $\partial /\partial x^1$ each
other, as $R_{,0} \rightarrow R_{,1}$ and $R_{,1} \rightarrow R_{,0}$.   
Other components such as $R_{(0102)}, R_{(0112)}$ etc vanish.

The tetrad components of the Ricci tensor are given by $R_{(\gamma
\delta)} \equiv \zeta^{\alpha \beta} R_{(\alpha \gamma \beta \delta)}$ 
and their non-zero components are
\begin{equation}
 \label{eq:A9}
{1 \over 2} (R_{(22)}+ R_{(33)}) = e^{-2a} R_{,01}/R = 0,
\end{equation}
\begin{equation}
 \label{eq:A10}
{1 \over 2} (R_{(22)}- R_{(33)}) = e^{-2a}  \Big[2 \gamma_{,01} +
(\gamma_{,0} R_{,1}+\gamma_{,1} R_{,0})/R  \Big] = 0,
\end{equation}
\begin{equation}
 \label{eq:A11}
R_{(00)} = e^{-2a} \Big[R_{,00}/R + 2 a_{,0}R_{,0}/R + {1 \over
2}(R_{,0}/R)^2 -2 (\gamma_{,0})^2 \Big]= 0,
\end{equation}
\begin{equation}
 \label{eq:A12}
R_{(11)} = e^{-2a} \Big[R_{,11}/R + 2 a_{,1}R_{,1}/R + {1 \over
2}(R_{,1}/R)^2 -2 (\gamma_{,1})^2 \Big]= 0,
\end{equation}
\begin{equation}
 \label{eq:A13}
R_{(01)} = e^{-2a} \Big[R_{,01}/R - 2 a_{,01} + {1 \over
2}(R_{,1} R_{,0})/R^2 -2 \gamma_{,0} \gamma_{,1}\Big] = 0.
\end{equation}

From Eqs.(A11) and (A12), the equations for $a_{,0}$ and $a_{,1}$
derived and it is shown that their differentiations with respect 
to $x^1$ and $x^0$, respectively  give the same expression for
$a_{,01}$ consistent with Eq. (A13), when we use Eqs. (A9) and
(A10). Accordingly Eqs.(A11) and (A12) are integrable for $a$, as
in Eq. (\ref{eq:ba5}) in the text.

If we use the Einstein equations (A9) $\sim$ (A13) and the
conditions $R_{,0} = const$ and $R_{,1} = const$, we obtain
\begin{equation}
 \label{eq:A14}
R_{(0101)} = - R_{(2323)} = e^{-2a} \Big[-{1 \over 2} R_{,0}R_{,1}/R^2 +2
\gamma_{,0} \gamma_{,1}\Big],  
\end{equation}
\begin{equation}
 \label{eq:A15}
R_{(0202)} = - R_{(0303)} = e^{-2a} (R_{,0})^2  \Big[-{3 \over 2R}
\gamma_{,0}/R_{,0} + 2 R (\gamma_{,0}/R_{,0})^3 - \gamma_{,00}/(R_{,0})^2 \Big],
\end{equation}
\begin{equation}
 \label{eq:A16}
R_{(1212)} = - R_{(1313)} = e^{-2a} (R_{,1})^2  \Big[-{3 \over 2R}
\gamma_{,1}/R_{,1} + 2 R (\gamma_{,1}/R_{,1})^3 - \gamma_{,11}/(R_{,1})^2 \Big],
\end{equation}
\begin{equation}
 \label{eq:A17}
R_{(0212)} = - R_{(0313)} = - {1 \over 2} e^{-2a}  \Big[{1 \over
2}R_{,0}R_{,1}/R^2 + 2 \gamma_{,0} \gamma_{,1} \Big].
\end{equation}
One of curvature invariants $R^{\alpha \beta \gamma \delta} R_{\alpha
\beta \gamma \delta}$ is proportional to
\begin{equation}
 \label{eq:A18}
{\cal{R}} = R_{(0202)}R_{(1212)} + (R_{(0212)})^2 +(R_{(0101)})^2
\end{equation}
Using Eqs.(A14) $\sim$ (A16) and Eq.(\ref{eq:ba4}) we obtain
\begin{equation}
 \label{eq:A19}
{\cal{R}} = R ~e^{-4 I} ~\Phi
\end{equation}
with 
\begin{eqnarray}
 \label{eq:A20}
\Phi &=&  \Big[\gamma_{,00}/(R_{,0})^2 -2R (\gamma_{,0}/R_{,0})^3 + 
{3 \over 2R}
\gamma_{,0}/R_{,0} \Big]  \Big[\gamma_{,11}/(R_{,1})^2 - 2 R
(\gamma_{,1}/R_{,1})^3 \cr 
&+& {3 \over 2R}\gamma_{,1}/R_{,1} \Big]
+ 2 \Big[- {1 \over 2R^2}
+ 2 \Big({\gamma_{,0} \gamma_{,1} \over
R_{,0}R_{,1}} \Big)^2 \Big]^2 
+   \Big[{1 \over 2R^2} + 2 \Big( {\gamma_{,0}
\gamma_{,1} \over R_{,0}R_{,1}}\Big)^2 \Big]^2. 
\end{eqnarray}
%

\section{HYPERGEOMETRIC FUNCTIONS}
Formula (\ref{eq:ba24}) for the hypergeometric
functions and their following expressions in series are found in the
texts on special functions [12].
\begin{equation}
 \label{eq:B1}
F(a,b,c;z) = 1 + \sum^\infty_{k=1} {a(a+1)\cdot\cdot\cdot
(a+k-1)b(b+1)\cdot\cdot\cdot (b+k-1) \over k!
~c(c+1)\cdot\cdot\cdot (c+k-1)}~ z^k,
\end{equation}
\begin{eqnarray}
 \label{eq:B2}
F_1(a,b,1;z) &=& \sum^\infty_{k=1} {a(a+1)\cdot\cdot\cdot
(a+k-1)b(b+1)\cdot\cdot\cdot (b+k-1) \over (k!)^2}\cr
&\times&
\Big[\sum^{k-1}_{l=0} \Big({1 \over a+l}+{1 \over b+l}
-{1 \over l+1}\Big) \Big]~ z^k,
\end{eqnarray}
\begin{eqnarray}
 \label{eq:B3}
F_2(a,b,c;z) &=& (-1)^c (c-1)! ~z^{k-1} \cr
&\times& \sum^{c-2}_{k=0}
{(-1)^k (c-k-2)! ~z^k \over (a-1)(a-2)\cdot\cdot\cdot (a-c+k+1)
(b-1)(b-2)\cdot\cdot\cdot (b-c+k+1)}\cr
&+&
\sum^\infty_{k=1} {a(a+1)\cdot\cdot\cdot
(a+k-1)b(b+1)\cdot\cdot\cdot (b+k-1) \over k! ~c(c+1)\cdot\cdot\cdot
(c+k-1)}\cr
&\times&
\Big[\sum^{k-1}_{l=0} \Big({1 \over a+l}+{1 \over b+l}
-{1 \over c+l}-{1 \over l+1}\Big) \Big] ~z^k,
\end{eqnarray}
\bigskip


\begin{figure}
\caption{Inhomogeneous vacuum models with a space-like singularity
at $R = 0$ and null singularities at $v = u = 0$. The null
 singularities are true only for $\alpha < 1/2$. Two figures are
 shown in case when ~(a) $R = v + u$ and ~(b) $R = -(v +u)$.
 \hfill\break}
\end{figure}
\begin{figure}
\caption{Inhomogeneous vacuum models with a time-like singularity
at $R = 0$ and null singularities at $v = u = 0$. The null
 singularities are true only for $\alpha < 1/2$. Two figures are
 shown in case when ~(a) $R = v - u$ and ~(b) $R = - v +u$.
 \hfill\break}
\end{figure}
\begin{figure}
\caption{Colliding gravitational waves with a space-like singularity
and null singularities. 
In regions I($v < v_0$ and $u < u_0$), II($v < v_0$ and $u > u_0$),
III($v > v_0$ and $u < u_0$), and IV($v > v_0$ and $u > u_0$), there
are the Minkowskian spacetime, the outgoing wave, the incoming wave,
and colliding waves, respectively.
These singularities are true only in the 
region IV.  
 \hfill\break}
\end{figure}
\begin{figure}
\caption{Colliding gravitational waves with a time-like singularity
and null singularities. These singularities are true only in the 
region IV in which $v > v_0$ and $u > u_0$. (a) Null singularities are 
in the lines be and bf. (b) The limiting case when one of null singularities
(bf) and a null boundary (ad) are overlapped. 
 \hfill\break}
\label{cfig4.eps}
\end{figure}

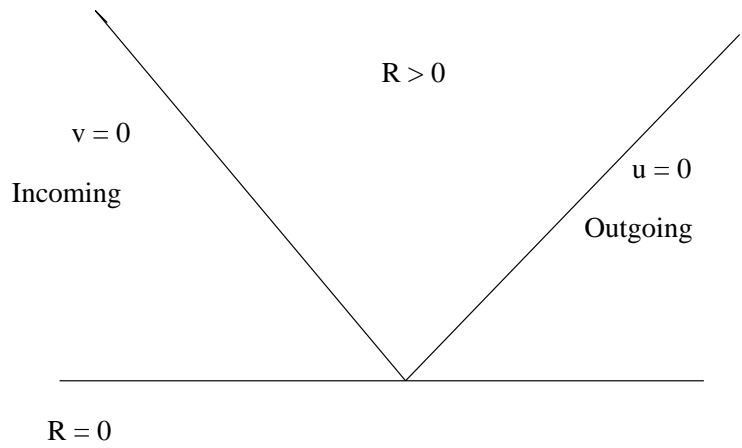
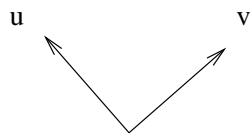
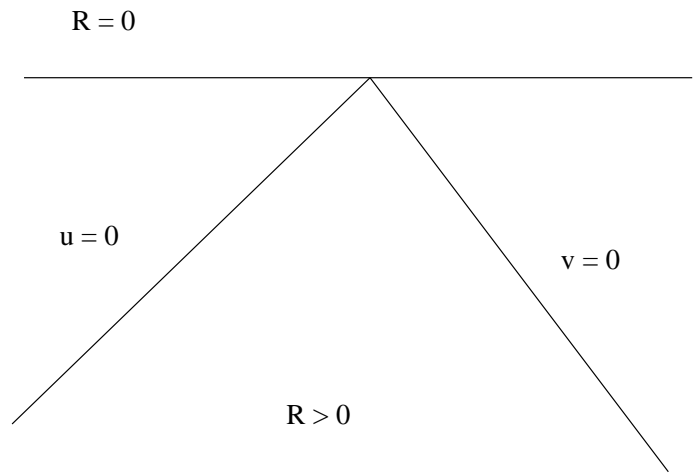
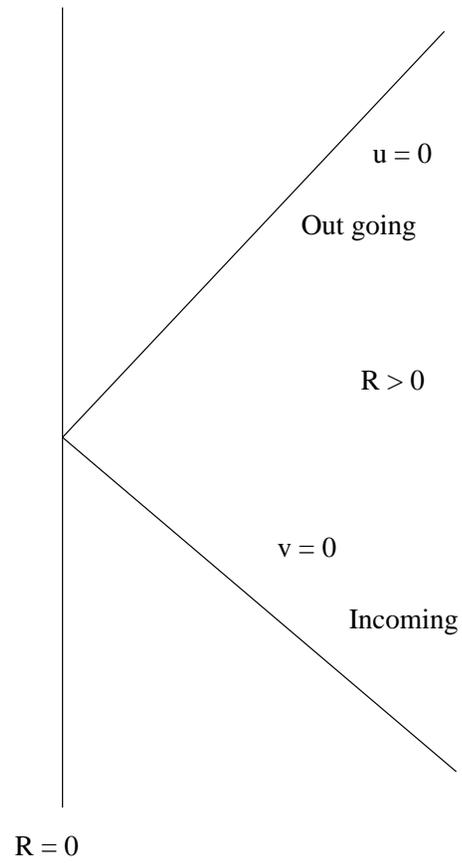
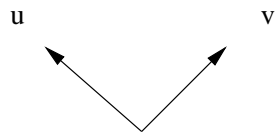
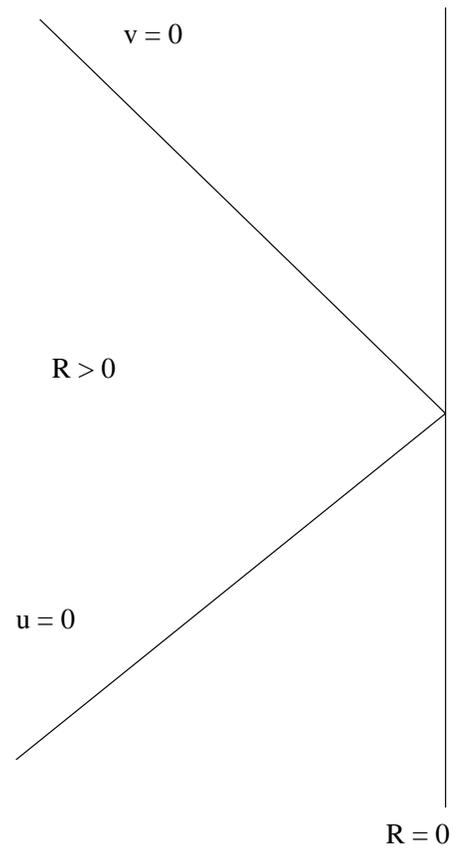
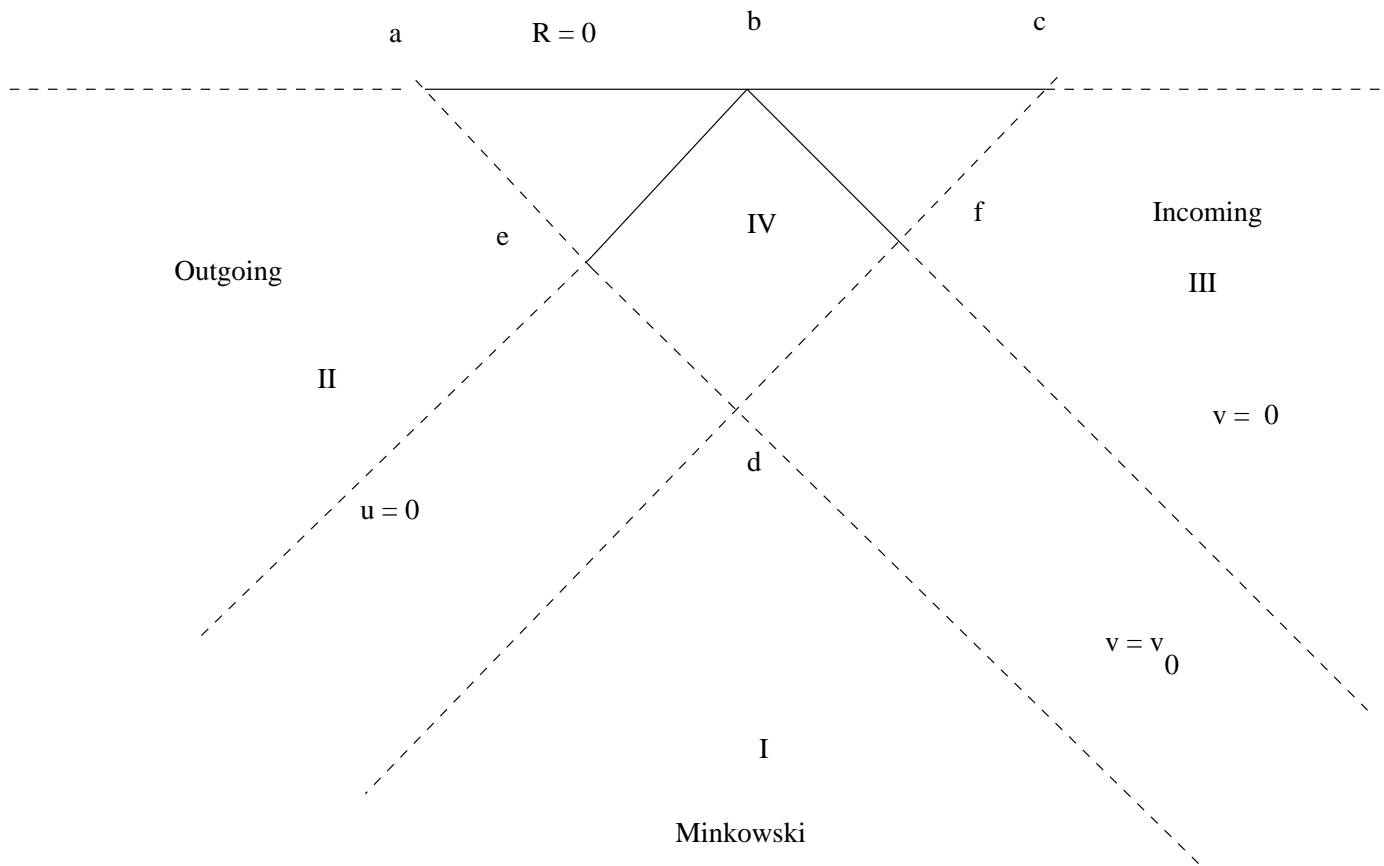
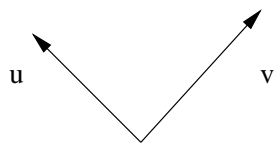
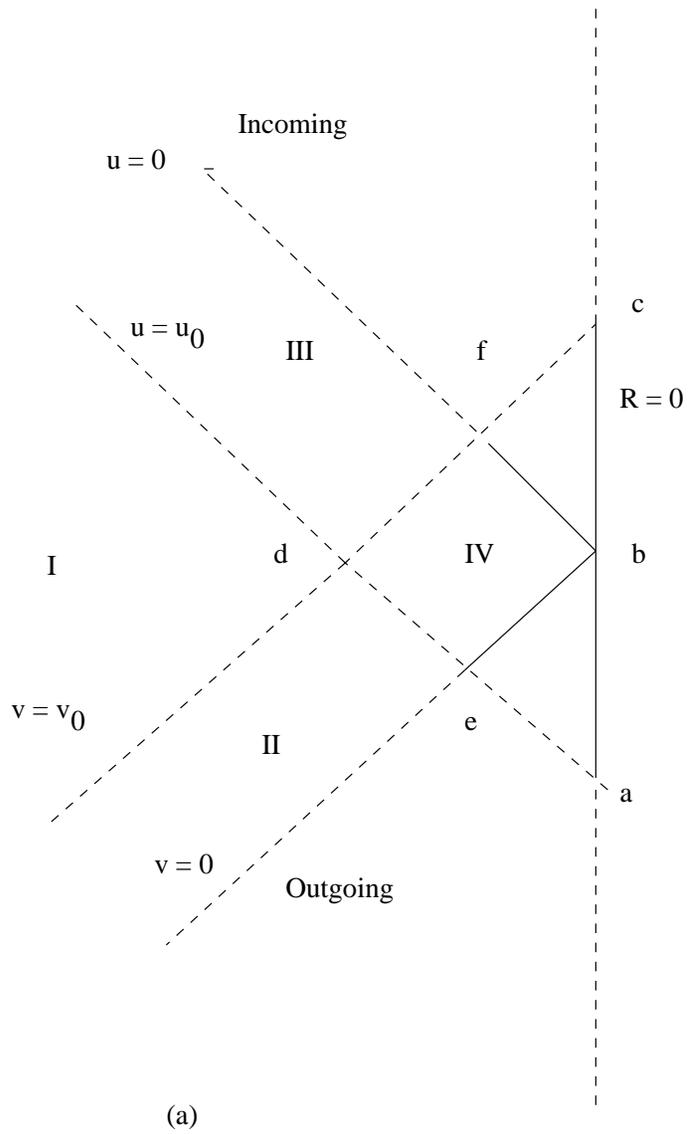
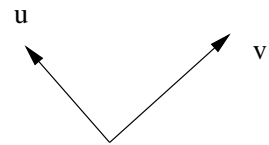
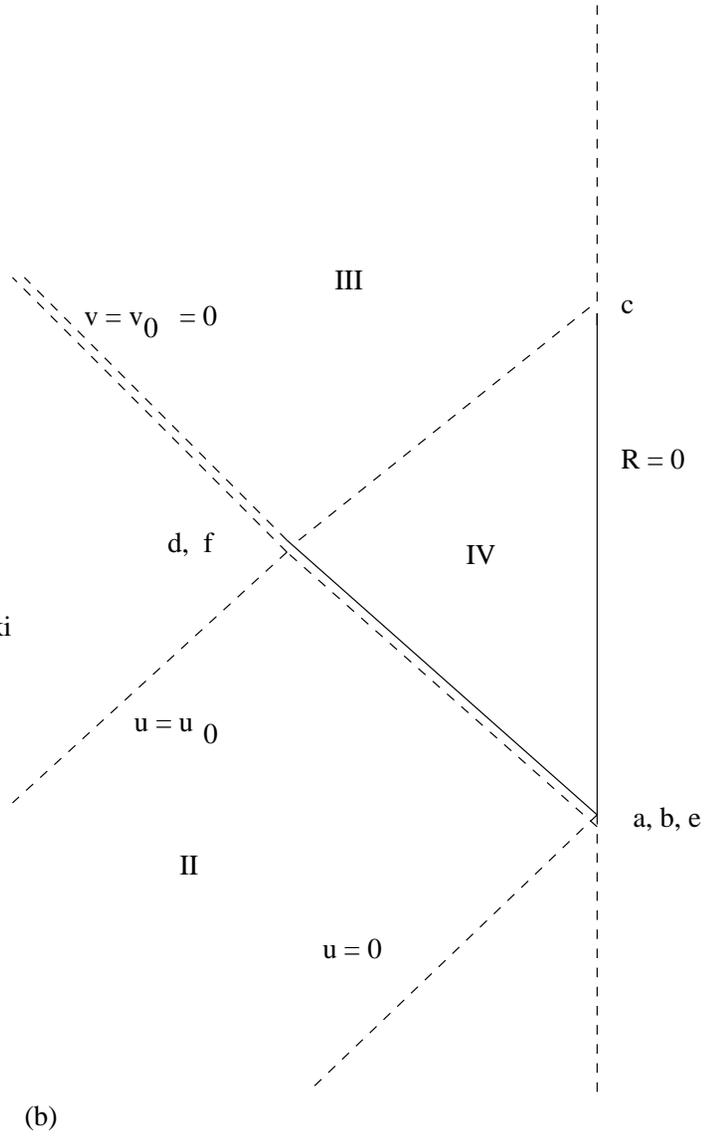
\end{document}